# A Fast Survey Focused on Methods for Classifying Anonymity Requirements

Morteza Yousefi Kharaji
Department of Computer Science and Information Technology
Mazandaran University of Science and Technology
Mazanadan, Iran
Yousefi@ustmb.ac.ir

Fatemeh Salehi Rizi
Department of Computer Science and Information Technology
Sheikh Bahaei University of Isfahan
Isfahan, Iran
Salehi.Fatemeh@shbu.ac.ir

*Abstract*—Anonymity has become a significant issue in security field by recent advances in information technology and internet. The main objective of anonymity is hiding and concealing entities' privacy inside a system. Many methods and protocols have been proposed with different anonymity services to provide anonymity requirements in various fields until now. Each anonymity method or protocol is developed using particular approach. In this paper, first, accurate and perfect definitions of privacy and anonymity are presented then most important problems in anonymity field are investigated. Afterwards, the numbers of main anonymity protocols are described with necessary details. Finally, all findings are concluded and some more future perspectives are discussed.

I. INTRODUCTION

Utilization of computer networks has been raised in recent years especially internet has become the most famous computer network in all over the world. While we are sending email or talking to our family members through internet, a lot of data or information packets are sent through internet. These packets consist of information of sender and receiver and etc. Since the packets are transmitted by several hops, everybody can monitor them and access to various information such as who started the contact and with whom and some other useful information. Although it is possible to conceal packet contents from a viewer by cryptography, the information of IP header is still accessible for a viewer. For this reason, in two past decades, some improvements have been emerged about anonymity and preserving privacy in formal and public communication field. So far, several systems have been designed and such systems are using by military groups, journalist and public sections. These systems are used to hide identities in virtual internet world .Today, there are various applications which need some methods to provide anonymity for performing their particular tasks. Some examples of these applications could be electronic voting, electronic commerce and etc.[1] Anonymity can be a branch of preserving privacy but preserving privacy is a concept wider than keeping anonymity of entities. Anonymous communication give a possibility to have contacts without disclosing their identities and it does not contain all aspects of privacy. Indeed, anonymity try to conceal operation agents' information while preserving privacy also hides whatever they perform [1].

Ignoring anonymity aspects causes to jeopardize people privacy. Hence, anonymity is one the most important issues in information security and preserving people privacy. So many applications need anonymity practically. In [3, 4, 5] these applications were categorized as follows:

- Searching information anonymously
- Maintain communication patterns to prevent traffic analysis
- Providing freedom of speech in fanatic environments
- Electronic voting
- Anonymous using of location based services
- Electronic payments
- Electronic cash
- Anonymous web browsing
- Anonymous e-mail
- Anonymous publishing

Anonymity attributes and also the level of anonymity are different in various applications. Therefore, analyzing of anonymity requirements which are used for determining accurate anonymity features in a service are very important and they must be done with high accuracy. For instance, applying a complete level of anonymity is not mostly a best choice because it causes some problems in many systems. There is no ability to follow and pursue entities in a complete anonymous system while the capability of imputing operations and attacks to people or entities in the system gives a possibility to hamper people's wrongdoings [2]. Consequently, anonymity must be applied with respect to the organization completely or under some particular conditions.

II. PRESERVING PRIVACY, ANONYMITY

Information is a lifeline in the most institutes, developed organizations and scientific communities. In the institutes and organizations in which information is really important, a quick and proper way is necessary to access to information. Organization and institutes should create informatics infrastructure and try to organize their information. One of success keys in institute, organizations and scientific communities in information age is speeding to generate and offer worthwhile information. After organizing information, it is necessary to provide regular and correct use of this information for others. Along with moving to developed organization based on information technology, it is essential to plan some other methods for preserving information.

Information security points to preserve information and minimize information revealing risks in unauthorized parts. Information security is a set of tools for preventing thefts, attacks, crimes, espionage and sabotages and etc. it is a science for studying various approaches to preserve data in computers and communication systems against access to unauthorized changes. Preserving privacy could be a subcategory of information security. Privacy means such a person can separate his/her information and disclose them on others view by his/her choice. Everyone has some private information which wants to keep them from others.

III. ANONYMITY ISSUES

Today, providing anonymity approaches are considered specially preserving entities' privacy in electronic commerce and electronic voting and etc. As it is mentioned before, content of messages could be protected by cryptography methods but message rout, source and destination of message, sending time, message length and some kind of information would still remain. Sometimes, valuable information can be accessible only by observing people communication pattern. Access to entities' information in a communication would be a violation of their privacy and anonymity can prevent revealing of this kind of information. Accordingly, anonymity can be a branch of information security [1].

Nowadays, there are a lot of applications that need anonymity and each application requires special anonymity attributes. For example imagine an electronic payment system that users can search their items and select and buy them. Most customers do not like to show their identity and their private information like interests and preferences. Thus, besides concealing users' identity connection between users' different operations must be hided. However, customers' anonymity should be applied in a limited way to preserve authority of trades correctly. It means that in electronic payment system, anonymity must be applied a different way. When an entity makes a wrongdoing, it would be possible to remove its anonymity and expose its real identity. As a matter of fact, the ability of imputing responsibility of operation to people gives a possibility to hinder crime activities in system by ordaining some rules and politics [2].

On the contrary, suppose an online medical consultation such that gives consultation to patients by hiding patients' identities. Since patients' backgrounds have very serious role in correct consultation, hiding information can damages system operation. Consequently, unlike electronic payment covering users' background might destroy accuracy of disease detection in a medical system.

Several protocols were proposed to provide anonymity in applications until now. It is necessary to have an organized method for developing software security because existence of this kind of method gives a capability to users for analyzing and describing application requirements. Therefore, it can reduce complexity of software analyzing and designing. Furthermore, it can save cost and time because it can recognize and move system problems in initial phase of software development.

VI. ANONYMITY PROTOCOLS

Anonymous communication means the communication layer must not reveal potentially identifiable information such as the user's IP address or location. This can be met by so-called anonymity protocols such as mix networks [6], onion-routing systems [7].

*A. Mix-Net Protocol*

The Mix-Net protocol is the base for some other anonymity protocols, Web Mixes [8], ISDN-Mixes [9], and Stop-and-Go-Mixes [10], to name a few. This protocol uses some nodes, called Mix, between sender and receiver. Mixes act as mediators for sending messages and provide the anonymity of the sender against the receiver. Moreover, Mixes are used for hiding a connection against attacks. Figure1 shows this protocol.

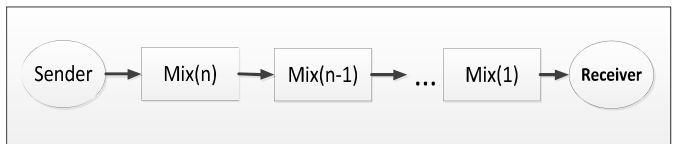

Figure 1: Mix-Net Protocol [11]

In the first step of executing the protocol, the sender defines a sequence of mixes. This can be accidentally or contractually. In this case it supposed the defined sequence is static. Then the sender encodes its message (M) by using the general key (Ki) of mixes. The sender adds the receivers address (AR) to the encoded message of last mix. The form of message is shown below:

Kn(Rn,(Kn-1(Rn-1, … , K2(R2, K1(R1, Ka(R0, M), AR))… )

In message form, there is a random number (Ri) besides of each encoded part. Therefore, before encoding the sender adds a set of random bits besides each part and this prevents the data from dictionary attack.

The important point in mix-net protocol is that even if one mix stay safe against traffic analyzing attack, the connection between sender and receiver will stay safe, because there is no relationship between the input and output of each mix [12, 13, 14].

### B. Onion Routing Protocol

In the Onion-Routing protocol, the sender and receiver can identify each other. The basic goal of this protocol is to make an anonymous connection from others' viewpoint, and to prevent network traffic analysis. This protocol uses a group of computers named Onion Routers. When a user has a request for sending a message, first the user considers a sequence of Onion Routers, then, makes a data for each router and uses layer encoding with general key encoding to preserve every router's data from other routers [15].

Each node peels a layer of onion, and this means the node decodes the information with its own private key that is related to itself and sends the result to next routers. After finishing this process of peeling a rout of onion routers is created between sender and receiver that can have an anonymous connection. According to this explanation the onion routing protocol creates a two-sided real-time connection between sender and receiver.

### V. PREVIOUSE WORKS ON CLASSIFICATION ANONYMITY REQUIREMENTS

These days, anonymity and preserving privacy are becoming very important issues in the digital world [5]. As a matter of fact, the requirements and the level of anonymity for different applications are different, and in many of the applications anonymity should be applied in a controlled and conditional manner. The concept of conditional privacy preservation has been widely studied in vehicular communications especially in VANETs [16]. The works in [17-20] are number of proposed methods to achieve conditional privacy.

Naessens et al. in [21, 22] introduced a methodology for designing controlled anonymity systems. This methodology defines four categories of requirements: Anonymity requirements, controlled requirements, applicability requirements and trust requirements. In their methodology, anonymity requirements come in a graph like "Unlinkable(X, Y)" which is called Linkability graph. In this system X and Y can be any kinds of operations or features. This graph shows for doing any operations what features needed to be accessed and what privileges will be required. Moreover, the proposed methodology uses Petri-Nets to represent the sequence of operations in a system. For each operation, it defines what kind of privileges will be required, when the operation will be finished, and what kind of privileges we will gain. The most important issue regarding this methodology is that it is not possible to consider all anonymity requirements from all aspects and put them into Unlinkable forms. Moreover, in this methodology there is no approach for detecting entities that might try to break the system rules.

Kavaki et al. in [23] proposed a methodology named Pris for considering privacy requirements in software development process. It is a Goal-Oriented methodology and defines the requirements as goals. The conceptual model that is used in Pris comes from Enterprise Knowledge Development framework that is shown in Figure 2. In this methodology, to reach the goals, they are divided into sub-goals until it is possible to reach each goal with a process. There exist several issues about this methodology; it divides all requirements (goals) into two categories: organizational goals and privacy goals which are too general. There are many applications with anonymity requirements, and these requirements are very different in each application; hence, sometimes considering the requirements in the form of such goals is not possible.

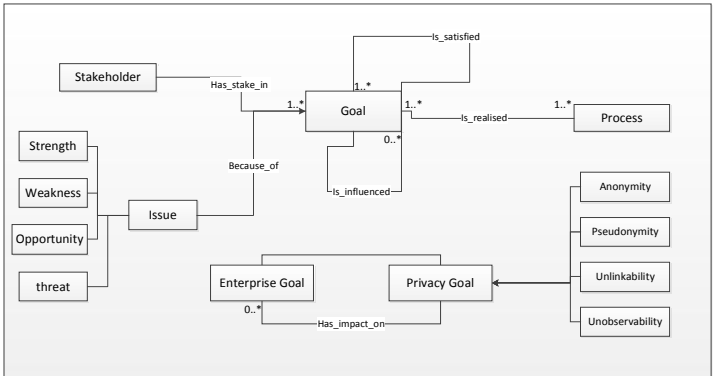

Figure2: PriS conceptual model [23]

As well as, Gürses et al in [24] proposed a methodology named CREE for Confidentiality Requirements Elicitation and Engineering which is applied to a real world project in the health care area. However, this work as stated by the authors is a primary effort and is also limited to the confidentiality and do not cover the anonymity concerns.

De Win et al. in [5] proposed a categorization for anonymity. In this categorization, they explained three features of anonymity such as traceability, linkability, and identifiability. They also proposed some combination of these features for any application that needs to be anonymous. For example if an entity is not traceable, linkable and identifiable, this entity is not anonymous, but if this entity is untraceable, unlinkable, and unidentifiable, it has the complete level of anonymity. In this approach the different combinations of these features make different levels of anonymity. Although this categorization is better than other works in this area but, this categorization is not complete enough, because they just consider some features of entities which mostly are related to the messages of entities or the connection between those entities. However, in this categorization they do not consider the features of entity itself which is selected to be anonymous.

TABLE I: DIFFERENT KINDS OF ANONYMITY [10]

| Traceability | Linkability | Identifiability | Anonymity |
|---|---|---|---|
| √ | √ | √ | 0(no Anonymity) |
| × | √ | √ | 1 |
| √ | × | √ | 2 |
| × | × | √ | 3 |
| √ | √ | × | 4 |
| × | √ | × | 5 |
| √ | × | × | 6 |
| × | × | × | 7 |

Spikerman and Cranor in [25] tried to offer a holistic view of privacy engineering and a systematic structure for the discipline's topics. They have used a three-layer model of user privacy concerns to relate them to system operations (i.e. data transfer, storage, and processing) and examine their effects on user behavior. Furthermore, they have developed guidelines for building privacy-friendly systems. An interesting result of [25] is that they have shown the degree of privacy friendliness of a system is inversely related to the degree of user data identifiability. However the levels of identifiability in [25] is limited to three levels: identified, pseudonymous, and anonymous.

VI. CONCLUSION

We live in electronic society and thus many of us read online news, manage online back account, buy online and chat with friends every day. Since we spend a lot of our daily time on the internet, anonymity treats are rising extremely. Storage memories are inexpensive; hence, the information of our activities can be saved and marinated with very low cost. Fortunately, a lot of efforts have been performed to preserve users' privacy and to anonymize users' communications in cyberspace up to now. The numbers of these existence anonymity protocols and methods with different approaches were studied in this paper. An accurate Knowledge of anonymity requirements in the system could be helpful to develop more secure and utilizable software and to have more safe online communication in the future.

REFRENCES